\documentclass[]{article}
\usepackage{emulateapj}
\usepackage{psfig}
\usepackage{graphicx}

\advance \voffset by 1.00cm\relax

\def\msun{{\rm ~M}_{\odot}}
\def\rsun{{\rm ~R}_{\odot}}

\def\mdot{\dot M}
\def\mpy{{\rm ~M}_{\odot} {\rm ~yr}^{-1}}

\begin{document}

\title{New constraints on SN Ia progenitor models}

 \author{Krzysztof Belczynski\altaffilmark{1,2}, 
 Tomasz Bulik\altaffilmark{3}, Ashley J. Ruiter\altaffilmark{1}}

 \affil{
     $^{1}$ New Mexico State University, Dept. of Astronomy,
        1320 Frenger Mall, Las Cruces, NM 88003\\
     $^{2}$ Tombaugh Fellow\\
     $^{3}$ Nicolaus Copernicus Astronomical Center,
       Bartycka 18, 00-716 Warszawa, Poland;\\   
     kbelczyn@nmsu.edu, bulik@camk.edu.pl, aruiter@nmsu.edu}

\begin{abstract} 
We use the {\tt StarTrack} population synthesis code to discuss potential 
progenitor models of SN Ia: single degenerate scenario, semi-detached 
double white dwarf binary and double degenerate scenario. 
Using the most recent calculations of accretion onto white dwarfs, we consider 
SN Ia explosions of various white dwarfs in binary systems.
We calculate the evolutionary delay times from the zero age main sequence to the 
explosion, and verify which scenarios can satisfy the constraints found by 
Strolger et al.~(2004), i.e.~a delay time of 2-4~Gyrs. 
It is found that the semi-detached double white dwarf binary model is 
inconsistent with observations. Only SN Ia progenitors in the double degenerate 
scenario are compatible with observations in all tested evolutionary models, 
with a characteristic delay time of $\sim 3$ Gyrs. If double degenerate
mergers are excluded as SN Ia progenitors, we find that the single degenerate
scenario may explain the observed delay times for models with low common
envelope efficiency. It is also noted that the delay time distributions within 
the various tested SN Ia scenarios vary significantly and potentially can be 
used as an additional constraint on progenitor models. 
\end{abstract}


\section{Introduction}

Supernovae Ia are used as precise distance indicators, and
the results of deep supernova searches (Schmidt et al. 1998; 
Perlmutter 1999; Riess et al. 2001; Tonry 2003) 
have led to the conclusion that the Universe expansion rate is accelerating.
This along with the WMAP results (Tegmark et al. 2004) provide evidence that 
the matter density in the Universe is $\Omega_{\rm M}\approx 0.3$, and
that the cosmological  constant density is $\Omega_\Lambda\approx 0.7$.
These results depend crucially on the assumption that SN Ia are standard 
candles. This assumption could be tested if the origins of SN Ia are 
recognized. Therefore, any new observational and theoretical constraints on 
their progenitors are very useful. 

The theoretical search for SN Ia progenitors is usually carried out with 
the population synthesis method and it was pioneered by Tutukov \& Yungelson 
(1981) followed by the number of other studies (e.g., Iben \& Tutukov 1984; 
Tornambe \& Matteucci 1987; Jorgensen et al. 1997; Yungelson \& Livio 2000;
Han \& Podsiadlowski 2004; Fedorova, Tutukov \& Yungelson 2004). 
There are still many open problems in the field. 
As long as it is accepted that SN Ia are caused by the disruption of a white
dwarf, the specific physical conditions leading to the explosion are widely 
discussed. Some important issues concern the white dwarf mass at which 
disruption followed by a SN can take place, accumulation rates on the white 
dwarf accretors, the outcome of double white dwarf mergers (SN Ia versus 
accretion induced collapse to neutron star) or the recently discussed issue of effects 
of rotation on white dwarf fate. Also, formation rates of different 
proposed SN Ia progenitors are not well constrained. The fact is that we lack 
an understanding of several evolutionary processes involved in the
specific formation scenarios (e.g., common envelope phase). 
We refer the reader to several reviews for a detailed discussion on an open issue of SN Ia
progenitors (e.g., Branch et al. 1995; Renzini 1996; 
Livio 2000). 

We consider three basic scenarios for SN Ia progenitors: single degenerate 
scenario (SDS, Whelan \& Iben 1973; Nomoto 1982), semi-detached double 
white dwarf binary (SWB, Solheim \& Yungelson 2004) model and double 
degenerate scenario (DDS, Webbink 1984; Iben \& Tutukov 1984). 
In the SDS scenario a white dwarf (WD) accumulates 
mass from a non-degenerate companion star. Nuclear reactions are ignited in a 
WD which has accreted a sufficient layer of material on its surface. At the 
time of ignition the WD can have a mass below the Chandrasekhar limit, or the 
accretion may push it over this limit, and the ignition takes place when the 
WD is already unstable. Binaries consisting of two WDs may lead, under favorable
conditions, to two qualitatively different progenitor SN Ia models.
Both models involve WD pairs which are close enough such that at 
some point in their evolution, the less-massive (larger) WD overflows its 
Roche lobe, initiating the mass transfer phase. 
The SWB scenario takes place  if mass transfer is dynamically stable and the 
pair evolves through the AM CVn stage (e.g., Nelemans et al.\,2001) and the system 
may produce SN Ia if (i) the accreting WD is pushed over the Chandrasekhar 
limit or (ii) the accreted material is ignited on the WD of 
sub-Chandrasekhar mass leading to the edge-on lit detonation. 
The DDS takes place  when the  mass transfer is dynamically unstable and the 
less massive WD is rapidly accreted onto the companion WD. If the total 
mass of the WDs exceeds the Chandrasekhar limit the accretor either goes through
accretion induced collapse and forms a neutron star or explodes in SN Ia (see 
Livio 2000 for review). In this work we {\em assume} that DDS results in a SN 
Ia explosion, in order to assess the model viability based on the time delay 
observations.
All of the above scenarios imply an evolutionary time delay between formation 
of the binary on the zero age main sequence (ZAMS) and the SN Ia explosion.

The recent results of the Hubble Higher $z$ Supernova Search yielded 
a number of detections with redshifts up to $z=1.6$. This allowed 
experimental determination of the delay time between 
the formation of a star on the ZAMS and supernova explosion (Strolger et al. 2004).
They used three trial functions describing the delay time distribution:
exponential, wide Gaussian and narrow Gaussian. 
Using the exponential function to describe the delay 
leads to a lower limit of 2.2-2.6 Gyrs, while the Gaussian fits lead
to a constraint on the delay in the range of 2.4-3.8 Gyrs, or 3.6-4.6 Gyrs
depending on the assumed star formation rate. In a separate study Gal-Yam \&
Maoz (2004) constrained the delay time to be longer than 1.7 Gyr.

In this paper we concentrate on calculation of the evolutionary delay times 
for different SN Ia progenitors, and compare them with the observational 
estimates of Strolger et al. (2004).

\section{Model Description}

The detailed description of the {\tt StarTrack} population synthesis code 
is presented in Belczynski et al. (2002; 2005, in preparation). In the
following we give an overview of the calculation scheme and summarize the 
most important features of the code. 

\subsection{Standard model}

Each model calculation is started with an instantaneous starburst in which
$10^6$  binaries are formed and then evolved through the next 15 Gyrs. The 
mass of the primary 
(the more massive component) is drawn from the IMF with a three-component
slope ($-1.3/-2.2/-2.7$ with the exponent changing at $0.5$ and $1.0 \msun$; 
Kroupa \& Weidner 2003) within the range $0.8-20 \msun$. The secondary mass 
is taken within the $0.08-20 \msun$ range through a flat mass ratio 
(secondary/primary) distribution. The distribution of orbital separations is 
taken to be flat in the logarithm ($\sim 1/a$) and we assume a thermal 
eccentricity distribution ($2e$).

The evolutionary calculations for stars (binary components) are based on 
modified formulae of Hurley et al.~(2000). The modifications, which are 
described in Belczynski et al. (2002) are not relevant for this study 
(e.g., black hole mass spectrum). For all our presented calculations we evolve 
stars with a solar metallicity ($Z=0.02$). 

Orbits of binary systems are allowed to change due to a number of 
physical processes. The recent magnetic braking law of Ivanova \& Taam 
(2003) is included, with the saturation for fastest spinning stars. 
For tidal interactions we use the Hut (1981) equations but we increase the
efficiency of tidal forces by an order of magnitude to recover the cutoff
periods for several open clusters (Mathieu et al. 1992). We use Peters (1964) 
equations to include orbital decay due to the emission of gravitational
radiation (GR). We also calculate the orbital expansion due to wind mass loss 
from binary components (Jeans mode mass loss). The wind mass loss rates
are taken from Hurley et al.~(2000), but extended to include winds for low-
and intermediate-mass main sequence stars (Nieuwenhuijzen \& de Jager 1990).  

The {\tt StarTrack} Roche lobe overflow (RLOF) treatment involves the detailed 
calculation of the mass transfer rates based on radius-mass exponents 
calculated both for the donor stars and their Roche lobes. The results of our 
calculations have been compared to a set of published detailed RLOF sequences 
(Wellstein, Langer \& Braun 2001; Tauris \& Savonije 1999; Dewi \& Pols 2003) 
as well as to calculations obtained with an updated stellar evolution code 
(Ivanova et al. 2003). Our approach to the mass transfer calculations allows 
for the possibility of (i) conservative versus non-conservative RLOF episodes 
(ii) thermally driven RLOF versus nuclear/magnetic braking/gravitational 
radiation losses driven RLOF and (iii) a separation of systems as persistent 
or transient, depending on whether the donor RLOF mass transfer rate lies 
below the critical rate for instability to develop in the accretion disk 
(adopted from Dubus et al. 1999 or  Menou, Perna \& Hernquist 2002 for 
different compositions of transferred material).
 
For WD, neutron star (NS) and black hole (BH) accretors during dynamically 
stable RLOF phases the 
accretion is limited to Eddington critical rate with the rest of the
transferred material lost from the system with the specific angular momentum
of the accretor. For all the other accretors, we assume that only half of
the transferred material is accreted (Meurs \& van den Heuvel 1989) and 
the rest is lost with the specific binary angular momentum (Podsiadlowski, 
Joss \& Hsu 1992).

\subsection{Common Envelope Phases}

{\em Standard Energy Balance Prescription}
If dynamical instability is encountered the binary may enter a common
envelope (CE) phase. 
We use the standard energy equations (Webbink 1984) to calculate the 
outcome of the CE phase
\begin{equation}
 \alpha_{\rm ce} \left( {G M_{\rm don}^f M_{\rm acc} \over 2 A_{\rm f}}
-
{G M_{\rm don}^i M_{\rm acc} \over 2 A_{\rm i}} \right) =
{G M_{\rm don}^i M_{\rm don,env} \over \lambda R_{\rm don,rl}}
 \label{ce}
\end{equation}
 where, $M_{\rm don}$ is the mass of the donor, $M_{\rm acc}$ is the 
mass of the accretor, $M_{\rm don,env}$ is the mass of the donor's 
envelope, $R_{\rm don,rl}$
is the Roche lobe radius of the donor, and the indices $i,f$ denote the
initial and final values, respectively. Parameter $\lambda$ describes the
central concentration of the giant (de Kool 1990; Dewi \& Tauris 2000).
The right hand side of equation~\ref{ce} expresses the binding energy of
the donor's envelope, the left hand side represents the difference
between the final and initial orbital energy, and $\alpha_{\rm ce}$ is
the CE efficiency with which orbital energy is used to unbind the
stellar envelope. If the calculated final binary orbit is too small to
accommodate the two stars then a merger occurs. In our calculations, we
combine $\alpha_{\rm ce}$ and $\lambda$ into one CE parameter, and for
our standard model, we assume that $\alpha_{\rm ce}\times\lambda = 1.0$.
If a compact object spirals in the common envelope it may
accrete significant amounts of material because of hyper-critical accretion 
(Blondin 1986; Chevalier 1993; Brown 1995). We have incorporated the
numerical scheme to include the effects of  hyper-critical accretion on NSs
and BHs in our standard CE prescription (for details see Belczynski et al.
2002). 
 
{\em Alternative Angular Momentum Prescription}
In addition to the standard prescription for the common envelope
evolution based on comparing the binding and orbital energies (Webbink 1984)
we investigate the alternative approach (Nelemans \& Tout 2005), based on
the non-conservative mass transfer analysis by Paczynski \& Ziolkowski
(1967),
with the assumption that the mass loss reduces the angular momentum in a
linear
way. This leads to reduction of the orbital separation
\begin{equation}
{A_f\over A_i}=\left( 1 -\gamma {M_{\rm don,env}\over M_{\rm tot}^i}\right)
{M_{\rm tot}^f\over M_{\rm tot}^i}
\left({M_{\rm don}^i M_{\rm acc}^i \over M_{\rm don}^f M_{\rm acc}^f}\right)^2
\end{equation}
where $M_{\rm don,env}$ is the mass of the lost envelope, $M_{\rm tot}^i$, 
$M_{\rm tot}^f$ are the total masses of the system before and after CE, and 
$\gamma$ is a scaling factor. We use $\gamma=1.5$ following Nelemans \& Tout
(2005). 

\subsection{Mass Accumulation on White Dwarfs}  

In the following we describe mass accumulation on white dwarf accretors 
during dynamically stable RLOF phases. If mass transfer in a binary system
containing a white dwarf and a non-degenerate companion is dynamically unstable, 
the system goes through a common envelope phase and we assume that the white 
dwarf does not accrete any matter. If dynamical instability is encountered
for a binary with two white dwarfs we assume a merger. If the total mass of the
two merging WDs is higher than 1.4 M$\odot$ we assume a SN Ia explosion, independent 
of what type of WDs are merging. We comment on relaxing that assumption in 
the last section.    

Accretion onto WDs may lead to a number of important phenomena, like nova or
Type Ia SN explosions or even to an accretion induced collapse (AIC)
of WD to a NS. In contrast to previous population synthesis studies, we 
incorporate the most recent results to estimate the accumulation efficiencies 
on WDs, which is crucial for the formation of SN Ia progenitors. 
In particular we consider accretion of matter of various
compositions onto different white dwarf types. We also include the
possibility that neutron star formation can occur via an AIC 
of a massive ONeMg white dwarf (Belczynski \& Taam 2004a).  
The effect of an optically thick wind from the white dwarf surface, which 
can stabilize the mass transfer in the system at high mass transfer rates 
is taken into account (see Kato \& Hachisu 1994; Hachisu, Kato, \& Nomoto 
1996, 1999). The accumulation rate of hydrogen-rich and helium-rich matter 
is taken from Hachisu et al. (1999) and Kato \& Hachisu (1999) 
respectively (see also Ivanova \& Taam 2004). For the direct accretion of 
helium or carbon/oxygen matter onto the ONeMg white dwarfs we make use of 
the work of Kawai, Saio \& Nomoto (1987) in determining the evolution of 
the accreting white dwarf.  

In the last few years several groups have initiated calculations of the 
effects of white dwarf rotation and spin-up due to the accretion (e.g., 
Piersanti et al. 2003; Uenishi, Nomoto \& Hachisu 2003; Saio \& Nomoto 2004; 
Yoon \& Langer 2005). Some 
interesting results (obtained with 1- or 2-D simulations) were presented, 
e.g. the possibility of WD reaching  super-Chandrasekhar mass, the possibly 
easier mass accumulation for SDS progenitors or avoidance of SN explosion
through edge-on lit detonation at sub-Chandrasekhar mass. These results are 
not yet incorporated into our model, but they may have important consequences 
on progenitor models if they are confirmed.   

{\em Accretion onto Helium white dwarf.} 
If the mass transfer rate $\mdot_{\rm don}$ from the H-rich donor is smaller than 
some critical value $\mdot_{\rm crit1}$, there are strong nova explosions 
on the surface of the accreting WD, and no material is accumulated.  The  
accumulation efficiency $\eta_{\rm acu}=0.0$, i.e. the entire transferred 
material is lost from the binary. If the $\mdot_{\rm don} > \mdot_{\rm 
crit1}$ then the material piles up on the WD and leads to an inspiral. For 
giant-like donors we evolve the system through CE to see if the system 
survives; for all other donors we call it a merger and halt binary 
evolution. The critical transfer rate is calculated from:
\begin{equation} 
\mdot_{\rm crit1} = l_0 M_{\rm acc}^{\lambda} (X*Q)^{-1} \mpy
\label{wd01}
\end{equation}
where, $Q=6\times 10^{18}\ {\rm erg\ g}^{-1}$ is an energy yield of Hydrogen
burning, $X$ is the Hydrogen content of accreted material, and $l_0$ and 
$\lambda$ are coefficients. For Population I stars (metallicity $Z>0.01$) 
the values are $X=0.7, l_0=1995262.3, \lambda=8$, while for Population 
II stars ($Z \leq 0.01$) we get $X=0.8, l_0=31622.8, \lambda=5$ (Ritter 
1999, see his eq. 10,12 and Table~2).  

If the mass transfer rate from the He-rich donor is higher than $\mdot_{\rm 
crit2} = 2\times 10^{-8} \mpy$ all the material is accumulated ($\eta_{\rm 
acu}=1.0$) until the accreted layer of material ignites in a helium 
shell flash at which point degeneracy is broken and a main sequence 
helium star is formed. Following the calculations of 
Saio \& Nomoto (1998) we estimate the maximum mass of the accreted shell at 
which flash occurs: 
\begin{equation}
\Delta M = \left\{ \begin{array}{ll}
  -7.8 \times 10^4 \mdot + 0.13 & \mdot < 1.64 \times 10^{-6}  \\
  0 {\rm (instantaneous\ flash)} & \mdot \geq 1.64 \times 10^{-6} 
\end{array}
\right.
\label{wd02}
\end{equation}
where $\mdot$ is expressed in $\mpy$.

The newly formed helium star may overfill its Roche lobe, in which case 
either a single helium star is formed (He WD companion), a helium contact 
binary is formed (helium main sequence companion) or the system goes through CE 
evolution (evolved helium star companion). 

For the lower than $\mdot_{\rm crit2}$ transfer rates, accumulation 
is also fully efficient ($\eta_{\rm acu}=1.0$). However, the SN Ia 
occurs at the sub-Chandrasekhar mass: 
\begin{equation}
M_{\rm SNIa} = -400 \mdot_{\rm don} + 1.34 \msun ,
\end{equation}
where $\mdot_{\rm don}$ is expressed in $\mpy$.
For mass transfer rates close to $\mdot_{\rm crit2}$, the above
extrapolations from the results of Hashimoto et al. (1986) yield 
masses smaller than the current mass of the accretor, and we 
assume instantaneous SN Ia explosion. We do not 
consider the accumulation of heavier elements since they could only 
originate from more massive WDs (e.g., CO or ONeMg WDs), which would have 
smaller radii and could not be donors to lighter He WDs.  

{\em Accretion onto Carbon/Oxygen white dwarf.}
We adopt the prescription from Ivanova \& Taam (2004). In the case of 
H-rich donors for the mass transfer rates lower than $10^{-11} \mpy$ 
there are strong nova explosions and no material is accumulated 
($\eta_{\rm acu}=0.0$). In the range $10^{-6} < \mdot_{\rm don} < 
10^{-11} \mpy$ we interpolate for  $\eta_{\rm acu}$ from Prialnik \& 
Kovetz (1995, see their Table 1). For the rates higher than $10^{-6} \mpy$ 
all transferred material burns into helium ($\eta_{\rm acu}=1.0$). 
Additionally we account for the effects of strong optically thick winds 
(Hachisu et al. 1999), which blow away any material transferred over the 
critical rate
\begin{equation}
\mdot_{\rm crit3} = 0.75 10^{-6} (M_{\rm acc}-0.4)  \mpy.
\end{equation}
This corresponds to $\eta_{\rm acu} = \mdot_{\rm crit3} / \mdot_{\rm don}$ 
for $\mdot_{\rm don} \geq \mdot_{\rm crit3}$. The accretor is allowed to 
increase mass up to $1.4 \msun$, and then explodes in a Chandrasekhar mass 
SN Ia. In the case of He-rich donors, if the mass transfer rate is higher than 
$1.259 \times 10^{-6} \mpy$ helium burning is stable and contributes to the 
accretor mass ($\eta_{\rm acu}=1.0$). For the rates in the range 
$5 \times 10^{-8} < \mdot_{\rm don} < 1.259 \times 10^{-6} \mpy$ accumulation 
is calculated from
\begin{equation}
\eta_{\rm acu} = -0.175 (\lg(\mdot_{\rm don})+5.35)^2+1.05
\end{equation}
and represents the amount of material that is left on the surface of
the accreting WD after the helium shell flash cycle (Kato \& Hachisu 1999). 
The mass of the CO WD accretor is allowed to increase up to $1.4 \msun$, and 
then a Chandrasekhar mass SN Ia takes place in the two above He-rich accretion 
regimes. If mass transfer rates drops below $5 \times 10^{-8} \mpy$, the 
helium accumulates on top of the CO WD and once the accumulated mass reaches 
$0.1 \msun$ (Kato \& Hachisu 1999), a detonation follows and ignites the CO 
core leading to the disruption of the accretor in a sub-Chandrasekhar mass SN Ia 
(e.g., Taam 1980; Garcia-Senz, Bravo \& Woosley 1999). If the mass of the  
accreting WD has reached $1.4 \msun$ before the accretion layer has reached $0.1 
\msun$ then the accretor explodes in a Chandrasekhar mass SN Ia. Carbon/Oxygen 
accumulation takes place without mass loss ($\eta_{\rm acu}=1.0$) and leads 
to SN Ia if Chandrasekhar mass is reached.  

{\em Accretion onto Oxygen/Neon/Magnesium white dwarf.}
Accumulation onto ONeMg WDs is treated the same way as for CO WD accretors.
The only difference arises when an accretor reaches Chandrasekhar mass.
In the case of ONeMg WD this leads to an accretion induced collapse and neutron
star formation, and binary evolution continues (see Belczynski \& Taam
2004a; Belczynski \& Taam 2004b).

\section{Results}

\subsection{Typical evolution leading to SN Ia}

The following examples of evolution leading to the formation of SN Ia
progenitors in three different scenarios were calculated within 
our standard evolutionary model (see \S\,2.1).

{\em DDS Example.} The evolution starts with two intermediate-mass 
main sequence stars ($M_1=6.0, 
M_2=4.5 \msun$) on a rather small ($a \sim 200 \rsun$) and  
eccentric orbit ($e \sim 0.7$). First RLOF begins at $\sim 70\ {\rm Myrs}$ 
right after the primary evolves off the main sequence and is crossing the Hertzsprung Gap. 
The orbit has been already circularized ($a \sim 60 \rsun$). 
The ensuing MT is dynamically stable but proceeds on the thermal 
timescale of the donor and is characterized by a high transfer rate 
($\sim 3 \times 10^{-4} \mpy$).
After the rapid evolution through the Gap, the donor starts climbing the red 
giant branch, where mass transfer continues but at slower rate driven 
by the nuclear evolution of donor ($\sim 10^{-7} \mpy$).   
Mass transfer stops when most of the donor envelope is exhausted. 
By this point the 
system has changed significantly; the primary becomes the less massive 
component ($M_1=1.0 \msun$), the secondary is rejuvenated ($M_1=7.0 
\msun$) by accreted material, while the orbit has expanded ($a \sim 380 
\rsun$) after mass ratio reversal.
The rest of the primary envelope is lost in a stellar wind during core 
Helium burning  and the primary becomes a low-mass naked helium star. 
The primary keeps evolving and at later stages expands and initiates 
the second RLOF ($88\ {\rm Myrs}$ since ZAMS) which ends up in the formation of a 
CO WD  ($M_1=0.9 \msun$) when the primary is stripped this time of its 
helium-rich envelope. Then the secondary evolves off the main sequence and after 
$93\ {\rm Myrs}$ since ZAMS fills its Roche lobe while on the red giant 
branch. This leads to CE phase and a helium star ($M_1=1.4 \msun$) 
is formed, this time out of the secondary, while the orbit shrinks 
significantly ($a \sim 3 \rsun$). The secondary evolves and eventually 
starts another RLOF, which stops after about $10\ {\rm Myrs}$. The primary CO 
WD becomes quite massive ($M_1=1.1$) due to accretion while the 
secondary losses most of its helium-rich envelope and becomes a CO 
WD ($M_2=0.8 \msun$). This is the second mass ratio reversal in the 
system; the primary once again is the more massive binary component. 
A double CO WD binary, with total mass exceeding Chandrasekhar 
mass ($M_1 + M_2 = 1.9 \msun$), is formed on a tight orbit 
($a = 2.7 \rsun$) after about $100\ {\rm Myrs}$ since binary formation. 
The following orbital decay takes $\sim 5\ {\rm Gyrs}$ leading to the 
merger of two WDs and a possible SN Ia explosion.

{\em SWB Example.} The evolution starts with two relatively low-mass 
main sequence stars ($M_1=1.7, 
M_2=1.5 \msun$) on a wide ($a \sim 1000 \rsun$) and highly eccentric 
orbit ($e \sim 0.9$). After about 2 Gyrs the primary star evolves 
off the main sequence and shortly begins climbing up the red giant branch. 
The orbit circularizes and in the process the system becomes tighter 
($a \sim 200 \rsun$). Eventually the primary overfills its Roche lobe,
leading to a first CE phase.  The primary losses its entire envelope and 
becomes a 
He WD ($M_1=0.4 \msun$), and the orbit contracts farther ($a \sim 10 \rsun$).   
The secondary follows the same path as the primary -- it evolves off the 
main sequence  
(in 3 Gyrs since ZAMS), becomes a red giant, overfills its Roche lobe and 
forms a second He WD ($M_2=0.2 \msun$) in a second CE event (orbit 
contracts to $a \sim 0.08 \rsun$). The most recently formed (and lower 
mass) WD must be close to filling its Roche lobe so the gravitational 
radiation and associated orbital decay brings the system to contact 
within a Hubble time. At first RLOF proceeds with a high (but dynamically
stable) mass transfer rate ($\sim 10^{-6} \mpy$) but soon 
it drops down ($\sim 10^{-8} \mpy$) to the regime when the material can
accumulate on the primary leading to the final explosion and disruption 
of the primary WD. At the moment of explosion, the primary WD has accreted 
about $0.1\msun$ and exploded at sub-Chandrasekhar mass ($M_1=0.5 \msun$).

{\em SDS Example.} The evolution starts with two intermediate-mass main
sequence stars 
($M_1=4.5, M_2=3.4 \msun$) on a very wide ($a \sim 4400 \rsun$) and highly 
eccentric orbit ($e \sim 0.8$). The primary evolves all the way to the 
late asymptotic giant branch before filling its Roche lobe. The orbit
circularizes before contact is reached ($a \sim 1500 \rsun$). 
The RLOF leads to first CE phase, orbital contraction ($a \sim 60 
\rsun$) and formation of a CO WD ($M_1=0.9 \msun$) at $\sim 160\ {\rm Myrs}$ 
since ZAMS. The secondary takes another $\sim 100\ {\rm Myrs}$ to evolve off 
the main sequence.
This time due to the smaller orbit size, the Roche lobe is encountered 
when the donor (secondary) is on the red giant branch. The second CE phase
ensues, leading to further orbital contraction ($a \sim 0.5 \rsun$), and 
the exposed core of the secondary (which is non-degenerate) forms a naked 
helium main sequence star ($M_2=0.5 \msun$). The orbit slowly decays owing to the 
tidal spin up of the secondary, and finally after synchronization is reached 
the third RLOF is encountered when the Helium star exceeds its Roche lobe 
($a \sim 0.4 \rsun$) $275$ Myrs since ZAMS.  This time mass transfer is
dynamically stable ($\sim 10^{-8} \mpy$), and helium-rich material is 
transferred and accumulated on the CO WD primary. After about $5$ Myrs of
accretion the layer accumulated on the WD explodes leading to primary
($M_1=1.0 \msun$) disruption in sub-Chandrasekhar SN Ia.

\subsection{Delay Times} 

The delay times can be contributed to three major processes: 
(i) evolution of stars to form two WDs (DDS, SWB), or to 
form a WD in RLOF system with non-degenerate companion (SDS);
(ii) orbital decay (GR) to bring a system to contact after 
formation of WD-WD binary (DDS, SWB); 
(iii) accumulation of sufficient amount of material on the 
WD surface to initiate SN Ia explosion (SWB, SDS).

The first contribution (i) is set by the mass of the
secondary star, since the evolution time depends very strongly on
initial mass and is longer for lower mass stars. These times are on 
average: $0.1$ Gyr (DDS, secondaries $\sim 3-8 \msun$), $0.4-12$ Gyr 
(SWB, secondaries $\sim 1-3 \msun$), and $\sim 0.5$ Gyr (SDS, 
secondaries $\sim 2-4 \msun$).

The second contribution (ii) is set by the GR timescale for a given
system, which depends very strongly on the orbital separation of two WDs.  
The orbital separation in turn is basically set by the CE efficiency. 
The majority of SN Ia progenitors evolve through one or two CE phases. 
The only significant changes to orbital separation are encountered during 
these phases (orbital contractions by factors of 10-100). Therefore, this 
part of delay is set by the CE efficiency and it may vary over a wide range, 
basically from zero to a Hubble time. 

The third contribution (iii) is set by the accumulation efficiencies
described in \S\,2.3. For a given type of system, we expect a specific 
mass transfer rate and a corresponding accumulation rate which sets the 
time of SN Ia explosion. The efficient accumulation happens only in  
a given mass transfer range, and this part of the delay is easily predicted.  
On average these times are around one Myr for SWB and several Myrs for SDS 
progenitors.  

Summarizing, we see that for DDS progenitors the delay time is set basically
by the GR orbital decay, for SWB systems both GR and evolutionary effects
play an important role, while for SDS systems the major contribution comes from 
evolutionary times. In general, the time needed for the accumulation on the WD
surface does not play a significant role.

\subsection{Statistics} 

It may be inferred from the evolutionary histories of SN Ia progenitors that 
the CE efficiency is a major factor influencing the SN Ia delay times.
Almost all progenitors evolve through at least one CE phase.  
The CE phase basically sets the timescale for GR orbital decay of DDS and SWB
systems. For SDS systems the CE orbit contraction sets the initial
conditions for a RLOF phase with corresponding mass transfer rate, which if
falls into a specific regime, may lead a system to SN Ia explosion. 
Unfortunately, the CE phase is not well constrained, therefore we will
perform the calculation of the delay times with different CE efficiencies
and treatments. 

{\em Standard Model.}
It is found that within the SDS, exploding WDs are: mainly CO WDs (82\%) and
a smaller number of He WDs (11\%), ONeMg WDs (4\%) and  hybrid (CO
core/He envelope) WDs (3\%) with  various types of donors.
In the SWB progenitor group the systems are: double He WD binaries (77\%),
CO WD -- hybrid WD systems (21\%), with the rest being different
combinations of He, CO, hybrid and ONeMg WDs.
Within the DDS class we find that most of the SN Ia progenitors are CO-CO WDs
(88\%), and the rest are binaries hosting one or two ONeMg WDs (12\%).
The relative numbers of potential SN Ia progenitors are: 13\% (SDS), 49\% (SWB)
and 38\% (DDS).
We note that DDS progenitors constitute the most uniform group among three
progenitor classes.
                                                                              
{\em Alternative CE Models.}
In the models with decreased CE efficiency we note a decrease in the number of
potential SN Ia progenitors by factors of 3 and 6, corresponding to models with 
$\alpha \lambda =0.3$ and $\gamma=1.5$, respectively. This is
expected since many potential progenitors evolve through a CE phase, and will
not survive this phase if the efficiency is smaller than in our standard
model. However, as argued by Nelemans \& Tout (2005), the decreased CE
efficiency may be needed to explain the observed population of double WDs. 
We also note that for these models DDS is more efficient in producing
SN Ia progenitors (70\%) as compared to the combined SDS and SWB classes (30\%).                                                                       

In all models DDS progenitors (by definition) explode at Chandrasekhar mass. 
It is found in all calculations that progenitors in the SWB class explode at 
sub-Chandrasekhar mass.  
For the SDS scenario in the standard model and the model with $\alpha \lambda =0.3$ it
is found that only a small fraction of WDs (6\%) explode at Chandrasekhar
mass with the majority of systems (94\%) exploding at sub-Chandrasekhar mass. 
Only for the model with the alternative CE prescription with $\gamma=1.5$
will the fraction
of SDS progenitors exploding at Chandrasekhar mass become significant
(37\%).

\subsection{Delay time distributions}

For each scenario leading to a potential formation of a SN Ia 
we note the time $T$ between the formation of the binary system on ZAMS 
and the explosion.
This gives us a distribution of time delays.  Ideally we would wish to 
compare such a distribution with the observed one. However, observations 
provide us
with one characteristic property of the delay distribution, i.e.~the 
typical delay time. Thus for each distribution we obtain, 
we calculate two timescales: the mean $t_{\rm av}$ and the median $t_{50}$ and 
use them for comparison. In order to better characterize the distributions
we also calculate the 25\% and 75\% quantiles
of the distribution: $t_{25}$, and $t_{75}$. We assume that the characteristic
timescale of a given model lies somewhere between $t_{\rm av}$ and $t_{50}$.
We list these characteristic times in Table 1.

We present the distribution of delay times within the standard model
in the left panel of Figure~1. 
In the case of SDS the distribution peaks between 0.4 and 0.8 Gyr, 
and has a weak tail extending to later times. 
The typical systems in this group consist 
of a CO WD accreting either from a He main sequence star or
hybrid WD, originating from binaries with intermediate mass components
($M_{\rm zams} \sim 2 - 4 \msun$).

In the case of the SWB the distribution is bimodal. The long delay time 
peak ($\sim 10$ Gyrs) corresponds to a majority of systems in SWB class: 
He WD - He WD binaries (77\%, see \S\,3.3), while the short delay time 
peak ($\sim 0.5$ Gyrs) corresponds to CO WD - hybrid WD systems (21\%).
The double He WDs descend from the low mass stars (secondaries $\sim 1 
\msun$) and therefore the long evolutionary time adds up to the overall 
delay time (see \S\,3.2). 
The CO WD - hybrid WD progenitors descend from more massive stars
(secondaries $\sim 3 \msun$) and the evolutionary time does not 
contribute significantly for the overall delay time. The truncation at 
long times is artificial, as we are not considering the evolution beyond 
the Hubble time.

The distribution of lifetimes in the case of DDS is nearly flat in $TdN/dT$,
and stretches from about 0.2 Gyr to the Hubble time. This distribution 
is similar to the one obtained by e.g. Jorgensen et al.~(1997) and Yungelson 
\& Livio (2000).

In the top rows of Table 1 we show the characteristic times 
for the standard model. In the case of the SDS both the average and the mean
delay time are below 1 Gyr. 
 For the SWB model the typical timescales are dominated by the 
 late end of the distribution and are above 7 Gyrs.
 In the case of the DDS the average
and the median delay time are in the 1.8-3.2 Gyr range. 
Comparison of these times with the delay times found by Strolger et al.~(2004)
argues strongly for the DDS as progenitors of SN Ia.

Table 1 also contains the characteristic  delay
times for the two alternative models we have considered, and we present these 
distributions in the middle and right panel of Figure 1. 
In the model with $\alpha \lambda=0.3$ the situation is similar to  
the standard model described above. The typical delay times 
in the SDS case of 0.7-1.8 Gyr are smaller than the Strolger et al. (2004)
result, while in the SWB case these times are much higher.
On the other hand the delay times in the DDS case of 2.9-4.3~Gyr 
are again in agreement with Strolger et al. (2004).

The results of the calculations with the alternative description
of the CE phase ($\gamma=1.5$) are shown in the right panel
of Figure 1 and the bottom rows of Table 1. In this case the characteristic times
in each scenario (SDS, SWB, DDS) are consistent with the range
2-4 Gyrs of Strolger et al.~(2004). We note however that 
in the case of SWB the delay time distribution is bimodal 
and has a minimum just in the range of 2-7 Gyrs. 
We note that  the SDS model can be tuned with a proper 
treatment of the CE phase to agree with the Strolger et al.~(2004)
timescale. On the other hand the delay time distribution
in the DDS model is consistent with the 2-4 Gyrs range 
irrespective of the CE phase description.

\section{Discussion and Summary}

We have  calculated the distributions 
of delay times between a burst of star formation and supernovae 
type Ia explosions with the use of the {\tt StarTrack} population synthesis 
code. We have considered three different models 
of the CE evolution, characterized by two values of 
$\alpha \lambda$ and $\gamma$. We find that the distribution of
delay times in the DDS scenario is approximately 
flat in  $TdN/dT$, while for the case of SDS and SWB models
it can be bimodal.  We characterized the shape 
of these distributions by the average and the median, and compare
these times with the typical delay range of 2-4 Gyr found by 
Strolger et al. (2004).
The characteristic values describing the distribution
of delay times in the DDS are consistent with 
the 2-4 Gyr value in each of the three evolutionary models.
In the case of SDS we find that one can tune the 
CE description such that the average and median of the 
delay time distribution fits in the 2-4 Gyrs range. 
The typical delay times in the SWB SN Ia progenitor class are above 7 
Gyr except for the $\gamma=1.5$ model. However, in this case the delay 
time distribution is bimodal and has a minimum for the values 
corresponding to the average or the median.

The estimated long delay times were noted to favor the DDS over 
SDS SN Ia progenitor model by Gal-Yam \& Maoz (2004), who used Yungelson 
\& Livio (2000) results showing rather long delay times for the DDS model, 
as opposed to shorter delay times for SDS progenitors.
On the contrary, Dahlen et al. (2004) noted that the long delay times 
give support to SDS SN Ia progenitor model, but the statement was not backed
up with any arguments or calculations.  
Our results show that in general DDS delay times are longer than for SDS
progenitors (see Table 1). However, we find that for the model with the 
alternative CE treatment the SDS delay times are as long as DDS times, 
and are consistent with observations.  
 
It was also noted (e.g., Hoflich et al. 1996; Nugent et al. 1997) that 
observed properties of SN Ia favor explosion of WD at Chandrasekhar mass,
while the sub-Chandrasekhar mass models do not explain even subluminous, 
red SN Ia. As described in \S\,3.3 only for the DDS scenario do we expect
Chandrasekhar mass SN Ia in all models. All SWB systems explode at
sub-Chandrasekhar mass. The majority of SDS systems explode also at
sub-Chandrasekhar mass (standard and $\alpha \lambda=0.3$ models) and only
for the model with the alternative CE treatment (with $\gamma=1.5$) it is found that
a significant fraction of SDS progenitors explode at Chandrasekhar mass.    
Therefore, the mass of an exploding WD points toward the DDS scenario, with
the possibility that the SDS scenario may provide possible Chandrasekhar mass
progenitors in the model with $\gamma=1.5$ CE treatment. The SWB scenario is
basically excluded as the potential SN Ia progenitor.

We conclude that the DDS is the preferred progenitor type of SN Ia based on 
the comparison of the delay distribution times with the observational 
analysis by Strolger et al.~(2004). Our calculations show that  within the 
DDS scenario the SN Ia progenitor population consists mainly of CO-CO WD 
binaries, hence the uniform properties of SN Ia are a natural result.
Finally, it is now known that such systems exist since Napiwotzki et al. 
(2004) found a double degenerate system with a total mass exceeding the 
Chandrasekhar mass and a 4 Gyr merger time.   
However, we note that the recent calculations (e.g., Timmes, 
Woosley \& Taam 1994; Saio \& Nomoto 1998) may indicate that massive WD 
mergers would lead to accretion induced collapse rather than to SN Ia.
If this is the case, then our results for SDS SN Ia progenitors could be 
shown to be consistent with the delay time observations, provided that 
the CE efficiency is rather low. There may be also some observational
evidence supporting SDS progenitor model. Ruiz-Lapuente et al. (2004) have
searched for possible companions to Tycho Brahe's 1572 supernova, and
concluded that the most probable companion was a G star. Also, Hamuy et 
al. (2003)  observed SN 2002ic and reported strong hydrogen emission from
circumstellar material, and interpreted it in context of the companion star 
being an asymptotic giant star which has created a bubble of H-rich material
around the progenitor through stellar wind mass loss. However, this
interpretation was countered by Livio \& Riess (2003) who presented the 
picture in which a DDS progenitor can also account for the H-rich 
circumstellar material. The H-rich gas would originate from a common 
envelope phase which would result in the merging of a white dwarf 
and the degenerate core of the companion followed by a supernova. 
Also the results indicating that massive WD mergers 
lead to an accretion induced collapse, may now be opposed with the
calculations including rapid WD rotation (e.g., Piersanti et al. 2003).
The rotation stabilizes the accretion and the massive merger may actually 
lead to SN Ia, lending support to the DDS scenario.
It is also possible that SN Ia originate from various progenitor classes, 
and further observations are needed to resolve the issue and guide
theoretical studies.    

The shape of the delay time distribution is different for the three progenitor 
scenarios (see Fig~1). The different shapes could be tested against
observations in order to get better constraints on progenitor classes, e.g. 
$dN/dT\propto T^{-1}$ for DDS.  We note that in order to assess the viability of 
the SDS or SWB one should attempt to model the supernova survey results
with a bimodal distribution of delay times.

\acknowledgements 
We would like to thank Tom Harrison and Ron Taam for comments on this
project and anonymous referee for very detailed and useful report.
We acknowledge support from KBN through grant PBZ-KBN-054/P03/2001. 
TB wishes to thank for the hospitality of the NMSU Department of Astronomy, 
while KB and TB acknowledge local support from Pike Lilley (Alamo).

\clearpage


\begin{deluxetable}{ccccc}
\tablewidth{200pt}
\tablecaption{SN Ia Delay Times [Gyr] }
\tablehead{ $\alpha \lambda=1.0$& $t_{\rm av}$ & $t_{25}$ & $t_{50}$ &
$t_{75}$ }
\startdata
SDS & 0.85 & 0.32 & 0.43 & 0.62 \\
SWB & 7.60 & 3.78 & 8.21 & 11.5 \\
DDS & 3.30 & 0.60 & 1.76 & 4.70 \\
\cutinhead{ $\alpha \lambda=0.3$} \\
SDS & 1.89 & 0.12 & 0.73 & 3.10 \\
SWB & 10.5 & 10.6 & 12.6 & 13.8 \\
DDS & 4.39 & 0.65 & 2.92 & 7.57 \\
\cutinhead{ $\gamma=1.5$} \\
SDS & 2.90 & 0.83 & 2.61 & 3.85 \\
SWB & 4.57 & 0.73 & 1.10 & 10.6 \\
DDS & 3.06 & 0.34 & 1.18 & 4.71 \\
\enddata
\label{numbers01}
\end{deluxetable}


\clearpage

\begin{figure}
\includegraphics[width=0.32\textwidth]{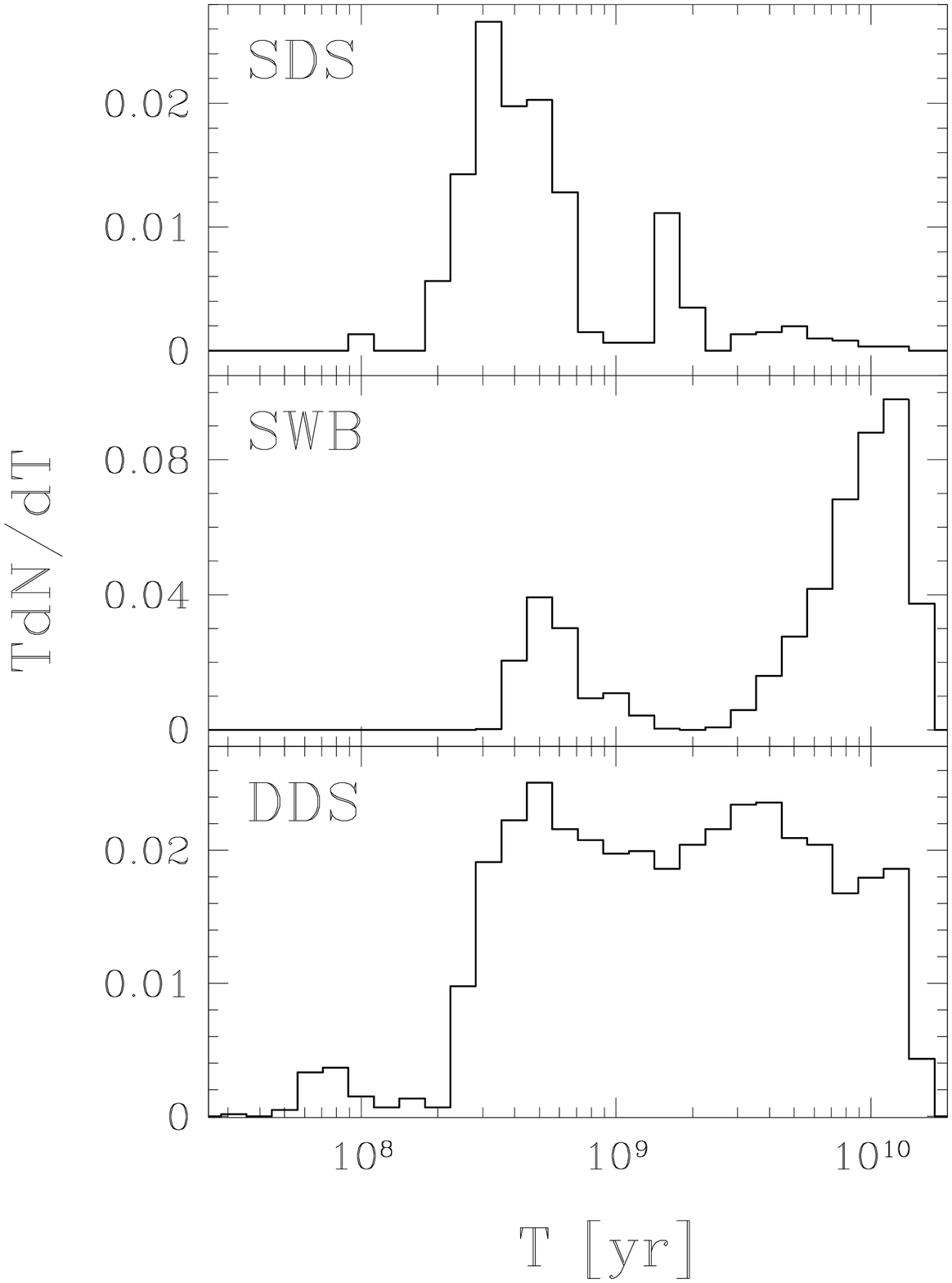}
\includegraphics[width=0.32\textwidth]{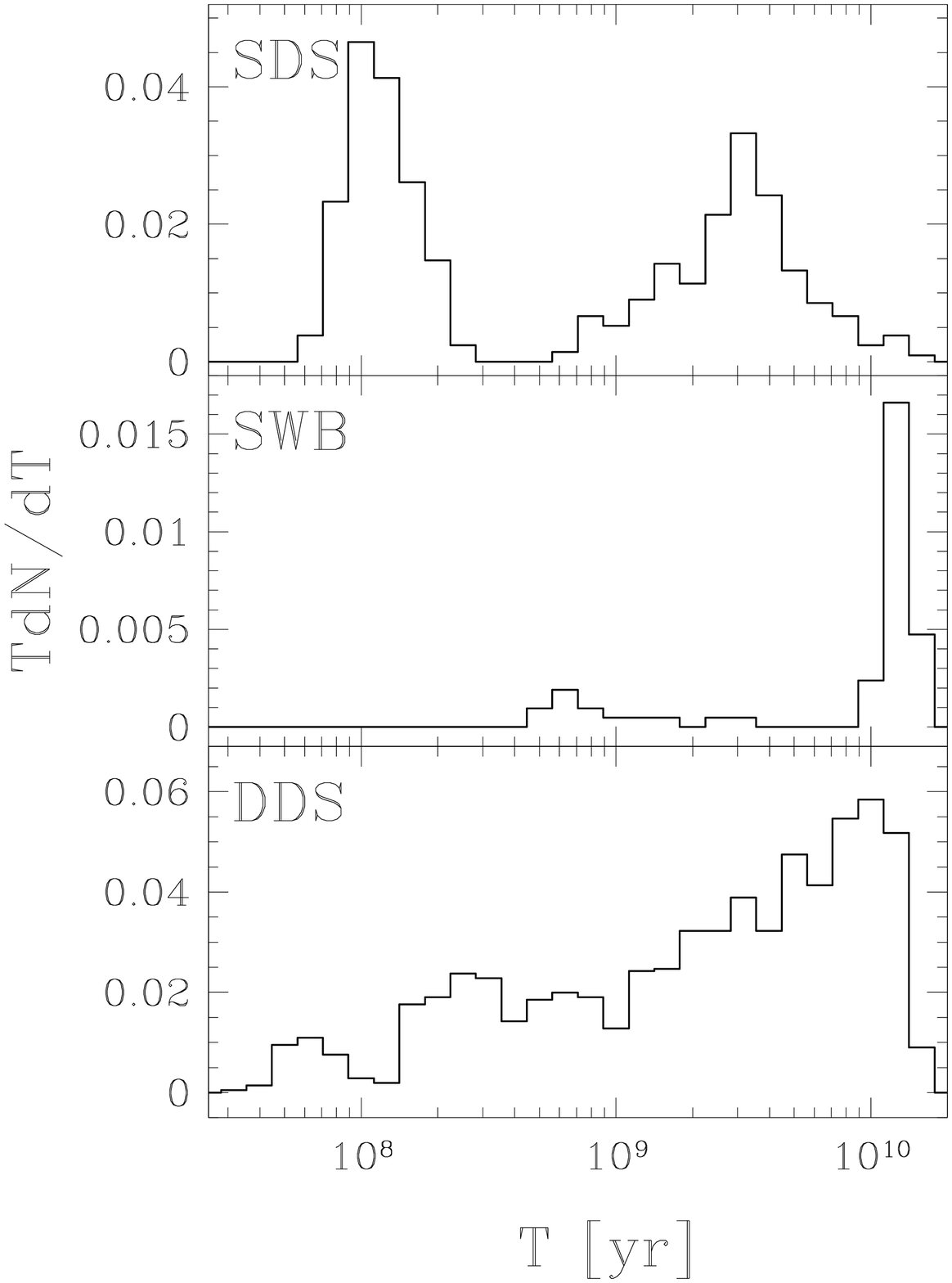}
\includegraphics[width=0.32\textwidth]{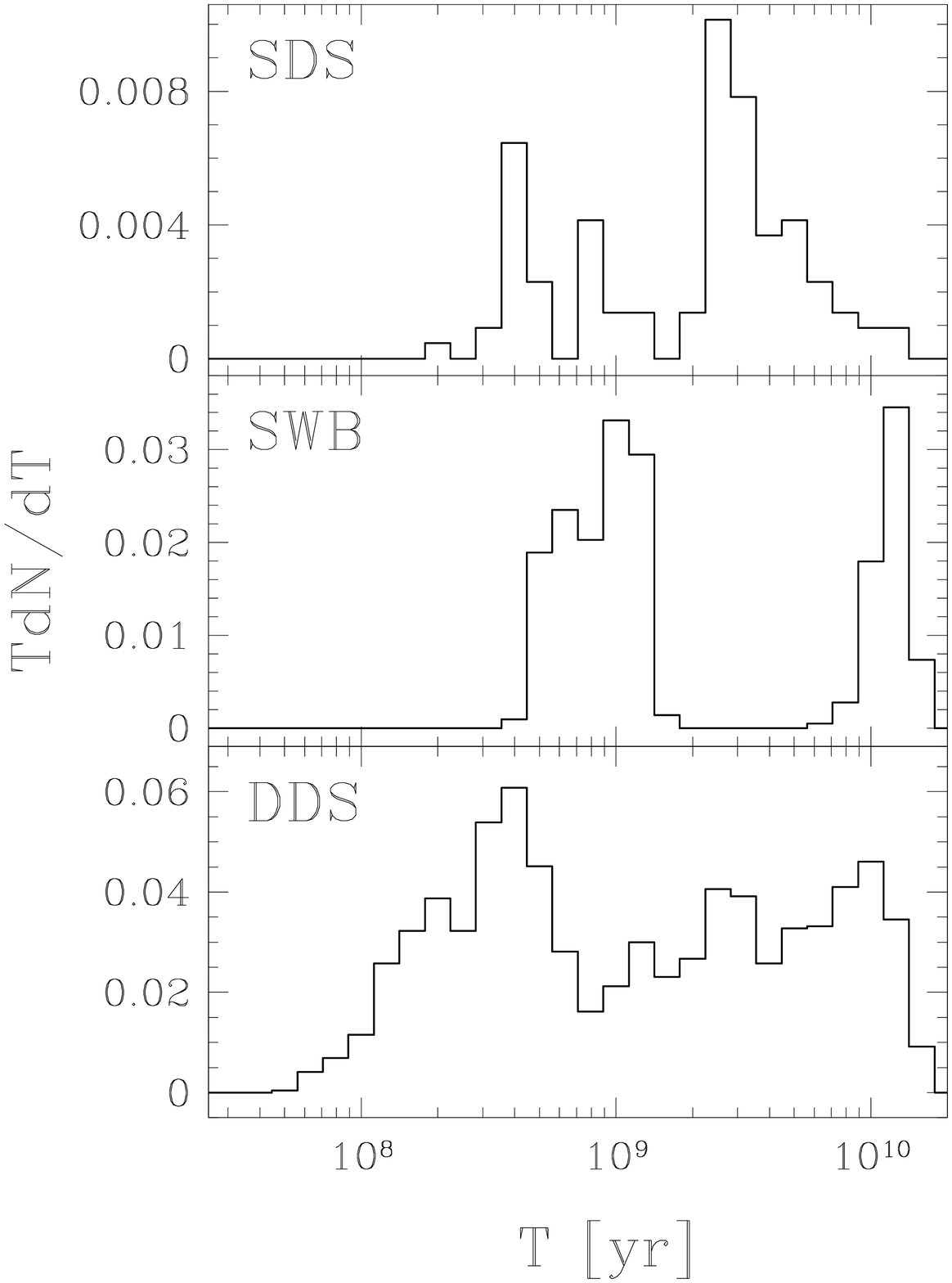}
\caption{
The distribution of SN Ia delay times since the burst of star formation.  
The left panel shows the results of our standard model calculation
($\alpha \lambda=1.0$), the middle panel corresponds to evolution
with the decreased CE efficiency ($\alpha \lambda=0.3$), while in 
the right panel we present results for alternative CE description
($\gamma=1.5$). The three different classes of SN Ia progenitors are shown on
each panel: single degenerate scenario (top), semi-detached double WD 
binaries (middle) and double degenerate scenario (bottom).  
The distributions are normalized to the sum of all SN Ia progenitors
(SDS+SWB+DDS) in the model. The data are binned in the intervals with the 
width of 0.1 in $\log(T/{\rm yr})$.}
\label{rys1}
\end{figure}


\end{document}